\documentclass[aps,prl,preprint,groupedaddress]{revtex4}
\usepackage{graphicx}
\usepackage{dcolumn}
\begin{document}

\title{Microcanonical Thermodynamic Properties of Helium Nanodroplets}

\author{Kevin K. Lehmann}
\email[]{Lehmann@princeton.edu}
\affiliation{Department of Chemistry, Princeton University, 
Princeton NJ 08544}

\date{\today}

\begin{abstract}
The density of states and other thermodynamic functions of helium
nanodroplets are calculated for a microcanonical ensemble with
both energy and total angular momentum treated as conserved quantum
numbers.  These functions allow angular momentum conserving evaporative cooling
simulations.
 As part of this project,  a recursion relationship is derived
for the reduction to irreducible representations of the n'th symmetric power of 
the irreducible representations of the rotation group.  These give
the distribution of total angular momentum states generated by putting
multiple quanta into a ripplon or phonon mode of the droplet, each of
which is characterized by a angular momentum quantum number.
\end{abstract}


\maketitle

\section{Introduction}

Spectroscopy of atoms and molecules dissolved in superfluid helium
nanodroplets has become a field of intense activity. 
Such nanodroplets rapidly cool by evaporation, reaching 
final temperatures of $\approx 0.37$\,K~\cite{Hartmann95}.
There is a long history of their formation and study by 
mass spectroscopic methods~\cite{Becker61,Northby01}.   
Vibrational~\cite{Goyal92,Hartmann95,Callegari01}, 
electronic~\cite{Stienkemeier95,Hartmann96a,Stienkemeier01}, and
rotational~\cite{Reinhard99,Callegari00c,Grebenev00a} 
transitions have been observed.  Part of the interest in
this field is that helium droplets provide a nearly ideal matrix~\cite{Lehmann98}
for isolation of highly unstable compounds. 
Many novel compounds and 
clusters have been
synthesized and spectroscopically studied in this unique 
environment~\cite{Goyal93,Hartmann96,Higgins96,Nauta99b,Nauta01c}.

There have been several previously reported calculations of the
evaporative cooling of helium nanodroplets~\cite{Gspann82,Brink90}, and these have 
predicted a terminal temperature in excellent agreement with
that later observed experimentally~\cite{Hartmann95}.  These calculations,
however, have not considered angular momentum conservation,
which obviously imposes constraints on droplet cooling in a
high vacuum environment.  
In order to address the question of the possible trapping of
angular momentum in a cooling droplet, we have undertaken a
new study of the evaporative cooling of the nanodroplets, but
including angular momentum conservation, using methods 
analogous to ``Phase Space Theory'' calculations of unimolecular
dissociation~\cite{Baer96}.  Necessary inputs to such calculations
are the density of states and integrated density of states of
helium nanodroplets, over the energy and angular momentum 
range sampled in the evaporative cooling trajectories of
nanodroplets. 

The angular momentum resolved density of states of helium nanodroplets is
not available in the literature.   While the calculation of
this quantity is largely based upon straightforward extensions
of standard convolution methods, there is one particular point that was not.
The distribution of angular momentum states produced by excitation of
an arbitrary number of quanta in a $(2L+1)$ degenerate vibrational 
state of angular momentum quantum number $L$ was required.  
This distribution is provided by the reduction of the $n$'th 
symmetric product of the $L$'th irreducible representation of
the group of transformations of a sphere, $K$.  While the equivalent
reduction for common molecular point groups is well known~\cite{Wilson55},
the present author could not locate the general result for $K$,
and has derived a recursion relationship that provides the needed 
reduction.   This is presented in the present paper, along with
a derived asymptotic expression, which is of the same form as
the thermal distribution of a rigid spherical top.  

The density of both ripplon and phonon modes are considered separately.
In each case, the simple scaling of the excitation spectrum with droplet
size (and thus number of helium atoms), allows the density of states
to be derived as a unique function of a reduced energy.  These
functions have been fit to simple analytical expressions which provide
excellent approximations over the range of energy and angular momentum
of interest.  In particular, it is found that the distribution of
angular momentum quantum states is of the form of the thermal 
distribution of rotation for a rigid spherical top, with a reduced energy
dependent effective inverse ``temperature'' that changes slowly with
energy.

\section{Ripplon Excitations}

The lowest energy excitations of a pure droplet are ripplons, which
are quantized capillary waves on the surface of the droplets.  
The properties of such waves on spherical drops are thoroughly
described in the classic text on hydrodynamics by Lamb~\cite{Lamb}.
Classically, each ripplon mode involves a modulation of the surface of the
droplet, with
\begin{equation}
r(\theta,\phi) = r_0 + \sum_{L,M} Re\left( A_{LM} Y_{LM} (\theta,\varphi) 
e^{i\omega_L t} \right)
\end{equation}
where $L \ge 2$ and
\begin{equation}
\omega_L = \sqrt{L(L-1)(L+2)} \, \omega_0  
\end{equation}
\begin{equation}
\omega_0 = \sqrt{\frac{\sigma}{m \rho R^3}} 
= \sqrt{\frac{4\pi \sigma}{3mN}}
\end{equation}
and $\sigma$ is the helium surface tension (approximated by the bulk
zero temperature value of 0.363 mJ/m$^2$), $m$ is the atomic mass of helium, 
$\rho $ is the number density of helium (approximated by
the bulk zero temperature value of 0.0218\,\AA$^{-3}$), and 
$R = \sqrt[3]{\frac{3N}{4\pi\rho}} = 2.22 N^{1/3}$\,\AA\
with $N$ equal to the number of helium
atoms in the droplet.
Given that the spectrum of ripplon modes is proportional to a common
factor that contains the size dependence, the thermodynamic functions of
the ripplon modes are universal functions of a reduced energy which
has unit value equal to 
$E_{\textrm{r}} = \hbar \omega_0 = 3.77/\sqrt{N}\,
k_{\textrm{\scriptsize{B}}}$K.

\begin{figure}[htbp]
\includegraphics[width=8in]{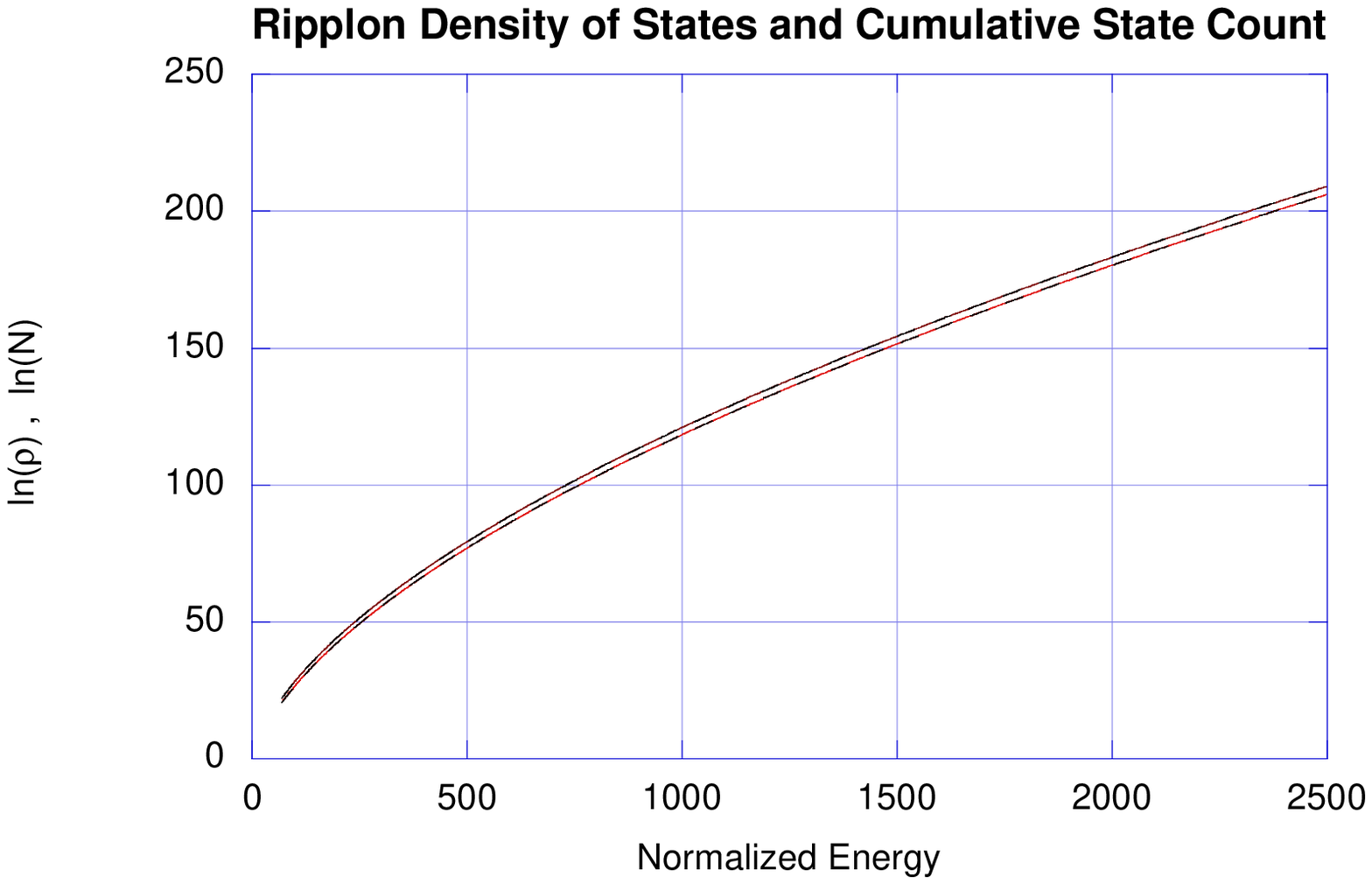}
  \caption{Plot of the natural log of the integrated density of states
and density of states in reduced energy units.  The integrated density
is slightly higher than the density of states.}
  \label{fig:rho_E}
\end{figure}

The integrated total state count, $N_E$, and the total density of states,
$\rho_E$, can be calculated from the spectrum of ripplon mode frequencies
using the Beyer-Swinehart direct count method~\cite{Baer96}.  In addition to
the spectrum, one needs the degeneracy of a state with $n_L$ quanta in
the ripplon mode with quantum number L.  This is equal to the number of
distinct ways of putting $n_L$ identical objects into $2L+1$ bins
and is equal to $g(n_L,L) = \frac{(n_L + 2L)!}{(2L)!n_L!}$~\cite{Reif65}.
Define $N_E(E,L)$ to be the integrated density of states calculated
using only ripplon modes with angular momentum quantum numbers
less than or equal to $L$.
We can recursively calculate the total
integrated density of states by using:
\begin{eqnarray}
N_E(E,L = 2) &=& \sum_{n_2} \Phi(E - \hbar \omega_2 n_2)
g(n_2,2) \\
N_E(E,L) &=& \sum_{n_L} N(E - \hbar \omega_L n_L, 
L -1) g(n_L,L) 
\end{eqnarray}
where the unit step function is defined by $\Phi(x) = 1(0)$ for x greater or equal to (or
less than) zero. In the Beyer-Swinehart algorithm, the above are calculated using
an energy bin size, $\delta E$ sufficiently small that each
$\omega_L/\delta E$ can be approximated by an integer.
A step size of $0.1$ reduced energy unit was used in this work.
The calculated values of $N_E$ and $\rho_E$ are plotted on figure~\ref{fig:rho_E}.
We have fitted the calculated integrated density of states 
to the expression
\begin{equation}
N_E = \exp \left( a E^{4/7} + b E^{1/7} \right) \label{eq:N_E}
\end{equation}
Values of a = 2.5118 and b= -3.4098 minimized the root mean squared
fractional errors  over the interval $E = 50--2500$.
Changing the step size in the calculation and refitting the
integrated density of states gave values of a = 2.5118 and b = -3.4110.
The leading power of $4/7$ can be derived by equating the high temperature
limit of the microcanonical and canonical ensemble values of energy and
entropy (which is proportional to $\ln(\rho_E)$ in the
microcanonical case).  The power of $1/7$ 
for the correction term was selected based upon the slope, on a 
log-log plot, of $N_E E^{-4/7}$ vs. $E$.  
The expression for $N_E$ given in Eq.~\ref{eq:N_E} agrees
with the Beyer-Swinehart direct count to within $\pm 5$\% 
while $N_E$ increases by more than 83 order of magnitude
over the reduced energy range $50-2500$.
Using standard expressions~\cite{McQuarrieText} for the
thermodynamic functions of a microcanonical ensemble, we can derive the following:
\begin{eqnarray}
\rho_E(E) &=& \left( \frac{{\textrm d}N_E}{{\textrm d}E} \right) =
N_E(E) \left[ \frac{4}{7} a E^{-3/7} + \frac{1}{7} b E^{-6/7} 
\right] \\
S_E(E) &=& k_{\textrm{\scriptsize{B}}} \ln (\rho_E(E)) 
= k_{\textrm{\scriptsize{B}}} \left[ a E^{4/7} + b E^{1/7}
 + \ln \left( \frac{4}{7} a E^{-3/7} + \frac{1}{7} b E^{-6/7} \right) \right]
\stackrel{E \gg 1}{\rightarrow} k_{\textrm{\scriptsize{B}}} a E^{4/7} \\
T_E(E) &=& \left( \frac{{\textrm d}S}{{\textrm d}E} \right)^{-1} =
\frac{1}{k_{\textrm{\scriptsize{B}}}} \frac{28 a E^{10/7} + 7 E}
{16 a^2 E + 8 a b E^{4/7} - 12 a E^{3/7} + b^2 E^{1/7}  - 6 b }
\stackrel{E \gg 1}{\rightarrow} \frac{1}{k_{\textrm{\scriptsize{B}}}} \frac{7}{4a} E^{3/7}
\\ C_V &=& \left( \frac{{\textrm d}T}{{\textrm d}E}     \right)^{-1} = \nonumber \\
&&\frac{ k_{\textrm{\scriptsize{B}}} \left( 16 a^2 E + 8 a b E^{4/7} - 12 a E^{2/7} - 6 b
\right)^2} {6 \left( 32 a^3 E^{10/7} + 32 a^2 b E - 56 a^2 E^{6/7} + 10 a b^2 E^{4/7}
-48 a b E^{3/7} + b^3 E^{1/7} - 7 b^2 \right) } \\
&  &  \stackrel{E \gg 1}{\rightarrow} k_{\textrm{\scriptsize{B}}} \frac{a}{4} E^{4/7}
\nonumber
\end{eqnarray}
$S_E$, $T_E$, and $C_V$ are the microcannonical values of
the entropy, temperature, and heat capacity due to the droplet modes
as a function of reduced rippon energy.
The density of states, $\rho_E$, can be compared with 
$0.311 E^{-5/7} \exp ( 2.49 E^{4/7} )$ derived by Brink
and Stringari~\cite{Brink90} through inverse Laplace Transformation
of the high temperature canonical partition function using the
stationary phase approximation. (Their reported results have
been converted to the normalized roton energy units used in this work).  It
can be seen that the exponential dependence of the density of
states is almost identical, but there is a slight difference in the
power of the energy dependence of the exponential prefactor.

For isolated helium nanodroplets, angular momentum as well as energy is a
conserved quantity, and thus the proper statistical ensemble is one that 
sums only over states of the same energy and total angular momentum.  
As a first step in the calculation of the properties of this ensemble, we
have derived a regression expression that determines $N_J(J,n_L,L)$ which
is the number of states of total angular momentum quantum
number $J$ that are generated from placing $n_L$ quanta in a
ripplon mode with angular momentum $L$.  We ``count'' such that
each ``state'' is an irreducible representation with $2J+1$
values of the projection quantum number.
This recursion expression and its
derivation is given in the Appendix to this paper. 

The individual values of $N_J(J,n_L,L)$ are quite irregular and do not appear
to derive from any simple expression.   However, the sum of states, of
course, gives the simple expression:
\begin{equation}
\sum_J (2J+1) N_J(J,n_L,L) = g(n_L,L) = \frac{(n+2L)!}{(2L)! n_L!}
\end{equation}
Also, the mean value of the squared angular momentum has been 
empirically found to be given by a simple expression:
\begin{equation}
< J(J+1) >_{n_L,L} = n_L \left[ L(L+1) + \frac{1}{2} L (n_L - 1) \right]
\label{eq:J2_nL}
\end{equation}
If we ignored the Bose symmetry (i.e. treated the angular momenta of different
quanta in the same Ripplon mode as uncorrelated), we would not have the last term
on the right hand side.   The Bose symmetry increases the mean squared angular
momentum because ripplon excitations have a greater probability of being in the
same direction than if they where distinguishable.   This closed form
expression can be used to sum a thermal distribution of ripplons as
a function of inverse temperature in reduced energy units, $\beta =
E_{\textrm{r}}/k_{\textrm{\scriptsize{B}}}T$.
\begin{eqnarray}
< J(J+1) > & = & \sum_{n_L,L} < J(J+1) >_{n_L,L} g(n_L,L) \exp 
\left( - \beta n_L \sqrt{L(L-1)(L+2)} \right) 
\nonumber \\ &=& 
\sum_L L(L+1)  \left[ \frac{2L+1}{e^{\beta \sqrt{L(L-1)(L+2)}} - 1} \right]
\left[ 1 + \frac{1}{2L \left( e^{\beta \sqrt{L(L-1)(L+2)}} - 1 \right)} \right]
\end{eqnarray}
If we neglect the correlations induced by the Bose symmetry of the ripplon
modes, we get the same result as above but with the last term on the right hand
side omitted.  

\begin{figure}[htbp]
\includegraphics[width=6in]{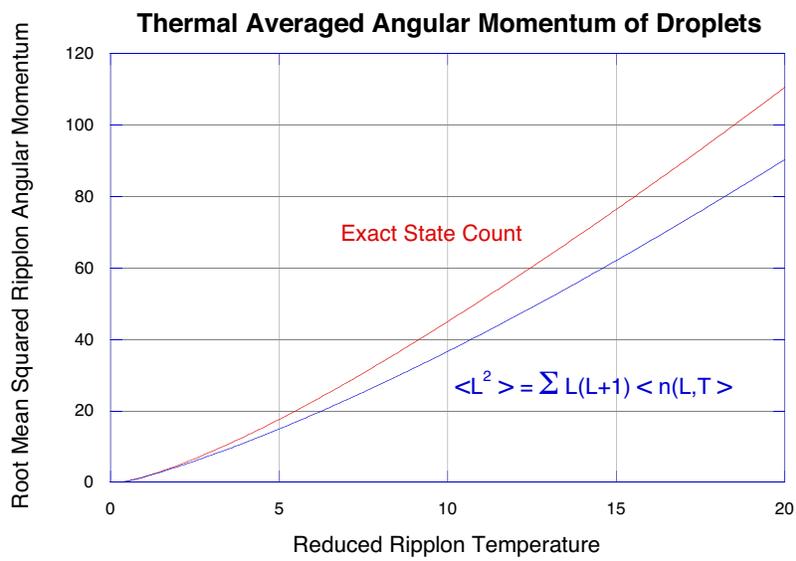}
  \caption{Plot of the Root Mean Squared total thermal angular momentum
in ripplons as a function of reduced temperature.  It can bee seen
that the effect of the Bose correlations of the ripplon excitations
increases the mean angular momentum by about 20\% over the 
uncorrelated result.}
  \label{fig:J2_plot}
\end{figure}

In a previous paper,the present author used the
uncorrelated expression to calculate the
root mean squared averaged angular ripplon angular momentum as
a function of droplet size~\cite{Lehmann99b}.  Figure~\ref{fig:J2_plot} shows a
plot of the thermally averaged value of ${\bf J}^2$ calculated both neglecting
the Bose symmetry and by explicit thermal average
over the present Microcanonical results.  It is seen that the exact
calculation is modestly higher than for the uncorrelated prediction,
but by a nearly constant factor of approximately 20\%.

\begin{figure}[htbp]
\includegraphics[width=6in]{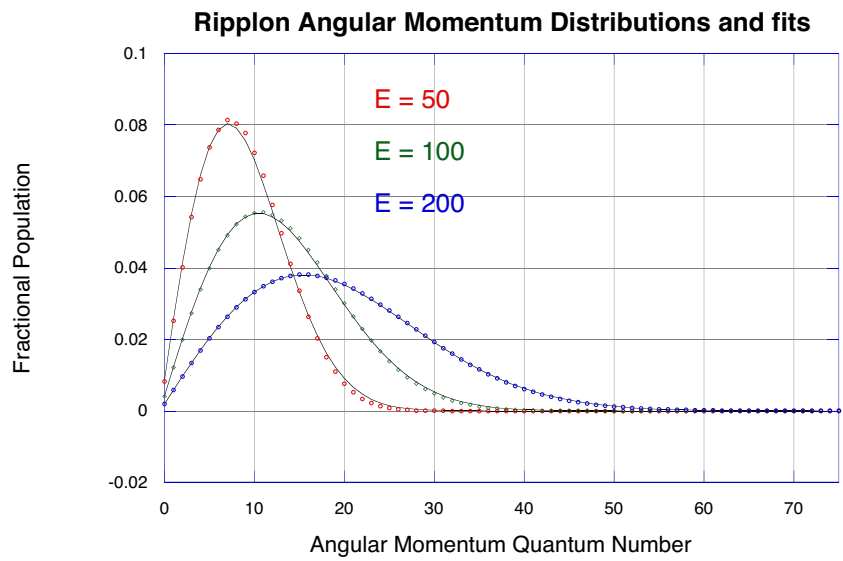}
  \caption{Plot of normalized integrated density of states as
function of total angular momentum quantum number for three
different values of the reduced energy.   Also plotted are
the distributions predicted by fits of these distributions to
a thermal distribution of a spherical top, as given
by Eq.~\ref{eq:N_EL}.}
  \label{fig:J_dist}
\end{figure}

Given $N_J(J,n_L,L)$, the
$J$ restricted state count and density of states can be computed using
a modification of the Beyer-Swinehart direct count.
Define $N_{EJ}(E,J,L)$ to the integrated density
of states with total angular momentum $J$, but
including excitation only in ripplon modes
up to angular momentum quantum number $L$. 
The following recursion relation is easily derived
by using the triangle rule:
\begin{eqnarray}
N_{EJ}(E,J, L = 2) &=& \sum_{n_2} \Phi(E - \hbar \omega_2 n_2)
N_J(J,n_2,2)  \\
N_{EJ}(E,J, L) &=& \sum_{n_L,J'} \sum_{J'' = |J - J'|}^{J+J'}
N_{EJ}(E - \hbar \omega_L n_L, J'', L -1) 
N_J(J',n_J,L)
\end{eqnarray}
for each $E$, we need to iterate on $L$ until $\hbar \omega_L > E$
to get the complete integrated density of states, $N_{EJ}(E,J)$,
with total angular momentum $J$.
We have explicitly calculated $N_{EJ}(E)$ for $E \le 200.$
by this procedure.
It was found that for each $E > 50$,  $N_{EJ}(E,J)$ has a distribution
that accurately fits the functional form:
\begin{equation}
N_{EL}(E,J) = N_E(E) (2J+1) \sqrt{\frac{\beta_L(E)^3}{\pi}} 
\exp \left( -\beta_L(E) (J + 1/2)^2 \right) \label{eq:N_EL}
\end{equation}
which is the form of the thermal distribution for a spherical rigid
rotor.   Figure~\ref{fig:J_dist} shows  plots of the calculated
distributions compared to the fitted ``thermal'' form
for several values of the reduced energy.

\begin{figure}[htbp]
\includegraphics[width=6in]{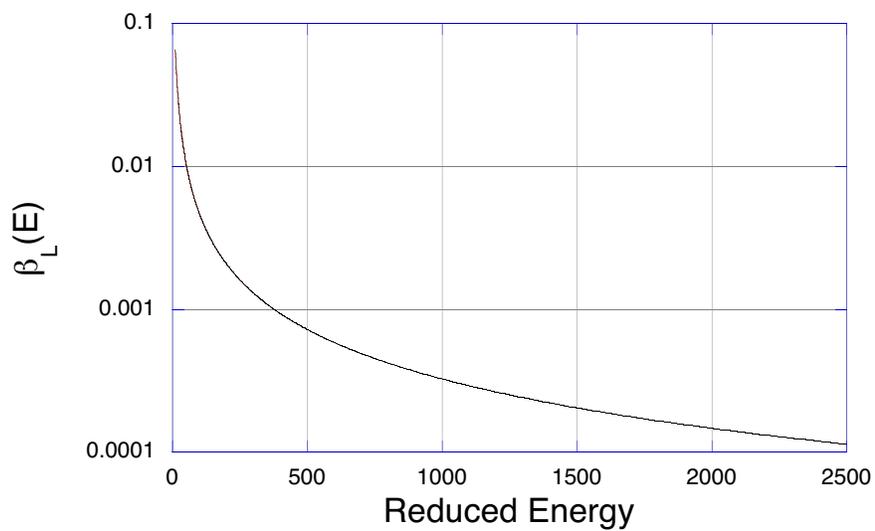}
  \caption{Plot of $\beta_L(E)$ as function of reduced energy.  
The values of $\beta_L$ where calculated both by from the
mean value of the squared total angular momentum, as calculated
by the recursion relations, and using Eq.~\ref{eq:beta_L}.  
These two values agree to better than the resolution of the 
plot.}
  \label{fig:beta_L}
\end{figure}

Assuming this ``thermal'' distribution of
states with different rotational quantum numbers, 
The value of $\beta_L(E)$ can be
calculated if we know the mean value of total
angular momentum, $<J(J+1)>$, over all states
up to a given energy by using $<J(J+1)> = \frac{3}{2} \beta_L^{-1}$.
Using the relationship given in Eq.~\ref{eq:J2_nL},
we have derived a recursion relationship to calculate
this quantity without having to explicitly enumerate all
the states as a function of $J$.
Let $A(E)$ be the average of the total angular momentum
quantum numbers for all states up to reduced energy $E$,
and $A(E,L)$ the same quantity, but calculated only using
ripplon modes with quantum numbers up to $L$.
Since angular momentum in different modes are uncorrelated, the
total angular momentum is just the sum of angular momentum in each individual
mode.   It is straightforward to derive that
\begin{eqnarray}
N_E(E,L) A(E,L) &=& \sum_{n_L} g(n_L,L) N_E(E - \hbar \omega_L n_L, L-1) \times
\nonumber \\
& & \left( A(E - \hbar \omega_L n_L,L-1) + n_L \left[ L(L+1) + 
\frac{1}{2} L (n_L - 1)
\right] \right) \label{eq:A}
\end{eqnarray}
The inverse ``rotational temperature'', $\beta_L(E)$, can be calculated by
using the standard spherical top relation $A(E) = 3 / (2 \beta_L(E))$.
Using this procedure, $\beta_L(E)$ has been calculated for $E = 10-2500.$.
The values of $\beta_L(E)$
are plotted in figure~\ref{fig:beta_L}
along with a fit through the points to a power law in
energy:
\begin{equation}
 \beta_L(E) = c E^{-8/7} + d E^{-13/7} \label{eq:beta_L}
\end{equation}
 with $c = 0.8680$ and $d = 0.9639$.  
A fit the the values calculated using an energy bin of 0.01 reduced
units gave $c = 0.8679$ and $0.9759$.
The $\beta_L$ values
from this fit are also displayed on figure~\ref{fig:beta_L},
but cannot be distinguished on the scale of the plot.
The $\beta_L(E)$ calculated with Eq.~\ref{eq:beta_L} agrees
with the calculated values to within 0.1\% for $E=100-2500$
and to within 1\%  for $E = 50-100$.  The reduction in accuracy
in this domain is because the calculated values of $\beta_L(E)$
oscillate around the value predicted by Eq.~\ref{eq:beta_L}. 
The leading power of -8/7 in $\beta_L(E)$ can be derived by
equating the high temperature limit for the mean
value of the total angular momentum squared for the
canonical and microcanonical ensemble.

Using the standard relationships, we can use calculate the thermodynamic
quantities as a function of reduced ripplon energy and total angular
momentum of the droplet as:
\begin{eqnarray}
\rho_{EJ}(E,J) &=& \left( \frac{\partial N_E}{\partial E}  \right)_J
= N_{EJ}  \left[ \frac{4}{7} a E^{-3/7} + \frac{1}{7} b E^{-6/7}
+ \left( \frac{3}{2\beta_L} - \left(J+\frac{1}{2} \right)^2   \right) 
\left( \frac{d\beta_L}{dE}   \right) \right] \\
S_{EJ}(E,J) &=& k_b \ln \left( \rho_{EJ} \right) 
= k_b \left[ a E^{4/7} + b E^{1/7} +\ln(2J+1) +\frac{1}{2} \ln \left(
\frac{\beta_L^3}{\pi} \right) - \beta_L \left( J+\frac{1}{2} \right)^2 
 \right. \nonumber \\
& & \left.  + \ln  \left[ \frac{4}{7} a E^{-3/7} + \frac{1}{7} b E^{-6/7}
+ \left( \frac{3}{2\beta_L} - \left(J+\frac{1}{2} \right)^2   \right) 
\left( \frac{d\beta_L}{dE}   \right) \right] \right]  \\
\frac{1}{k_b T_{EJ}(E,J)} &=& \frac{1}{k_b}
\left( \frac{\partial S_{SJ}}{\partial E} \right)_J  =
\frac{4}{7} a E^{-3/7} + \frac{1}{7} b E^{-6/7} + \left( \frac{3}{2\beta_L}
- \left( J + \frac{1}{2} \right)^2 \right) 
\left( \frac{d\beta_L}{dE} \right) \nonumber \\
& & + \frac{-\frac{12}{49} a E^{-10/7} -\frac{6}{49} b E^{-13/7} 
+ \left( \frac{3}{2\beta_L} - \left(J+\frac{1}{2} \right)^2   \right) 
\left( \frac{d^2\beta_L}{dE^2}   \right) 
- \frac{3}{2\beta_L^2} \left( \frac{d\beta_L}{dE}\right)^2 }
{\left[ \frac{4}{7} a E^{-3/7} + \frac{1}{7} b E^{-6/7}
+ \left( \frac{3}{2\beta_L} - \left(J+\frac{1}{2} \right)^2   \right) 
\left( \frac{d\beta_L}{dE}   \right) \right] } \label{eq:TEJ}
\end{eqnarray}

If we neglect  $\frac{3}{2\beta_L^2} \left( \frac{d\beta_L}{dE}\right)^2$
in Eq.~\ref{eq:TEJ} (which $\approx 3.2 E^{-4/7}$ times
 $\frac{12}{49} a E^{-10/7}$ that it is added to), then the equation for $T_{EJ}$
can be shown to reduce in the high energy limit to:
\begin{equation}
T_{EJ}(E,J) = T_E(E) + 0.48 E^{-9/7} \left( \frac{3}{2\beta_L}
- \left( J + \frac{1}{2} \right)^2 \right)  + O\left( \frac{3}{2\beta_L}
- \left( J + \frac{1}{2} \right)^2 \right)^2
\end{equation}
$T_{EJ}$ decreases monotonically with $J$ for fixed $E$.  
$T_{EJ}(E,0) = T_E(E) + 0.83 E^{-1/7}$,   
while for the RMS value of $J$, $T_{EJ}(E,J) = T_E(E)$.
However, for $J \gg 1.2 E^{6/7}$,
$T_{EJ} \rightarrow 1.01 E^{15/7} J^{-2}$, values much below 
$T_E(E)$ for the same value of $E$.  

\section{Phonon Excitations}

Starting at higher energy than the surface ripplon modes are phonon
(compressional) excitations of the helium droplets.   These normal modes
are characterized by two quantum numbers, $n, L$, where $n =1,2,\dots$
is the number of radial nodes and $L$ is the angular momentum quantum number.
Each normal node can be characterized by a wavenumber $k_{n,L}$ which is
determined by $k_{n,L} = r_{n,L}/R$ where $r_{n,L}$ is the $n$'th root
of the spherical bessel function $j_L$~\cite{Tamura96}.  For $k_{n,L} \ll 1$\,\AA$^{-1}$\
(i.e.  much smaller than the wavenumber of a roton), the excitation angular frequency
of each mode is given by 
$\omega_{n,L} =  u k_{n,L}$,
where $u = 236$\,m/s is the speed of sound in helium~\cite{Tamura96}.
We define a reduced energy $E_{\textrm{p}}$ with an energy unit
equal to the excitation energy of the lowest ($n=1, L=0$) phonon
which is $\hbar u \pi / R  = h \cdot 534 N^{-1/3}$\,GHz 
$= 25.5 N^{-1/3}$\,K\,$k_{\textrm{\scriptsize{B}}}$.

We have calculated the density of states using similar methods as 
for the ripplon density of states.
The integrated density of states was fit to the functional form
\begin{equation}
\ln(N_E) = a E^{3/4} + b E^{1/4} + f
\end{equation}
The values $a=3.3306$, $b=-3.5941$, and $f = 1.7786$ reproduces the 
recursively calculated values
of $N_E$ to an accuracy of better than $\pm 0.09\%$
while $N_E$ varies
by more than 130 orders of magnitude over the reduced
energy interval $E_{\textrm{p}} = 25-500$. 

We have assumed that the distribution of total angular momentum
states will again be well approximated by the  spherical top thermal distribution
expression~\ref{eq:N_EL}.
Values of $\beta_L(E)$ have been calculated using the recursion
relationship for $A(E)$, Eq.~\ref{eq:A}.
These were found to be well approximated
by the expression
\begin{equation}
\beta_L(E) = c E^{-5/4} + d E^{-7/4}
\end{equation}
Values of $c = 0.2545$ and $d = 0.2929$ reproduce the calculated
values of $\beta_L(E)$ to a fractional accuracy of better than $0.2\%$ over
the  reduced energy interval $E_{\textrm{p}} = 25-500$.  
Given these expressions for $N_E(E)$ and $\beta_L(E)$, expressions
for the various microcanonical thermodynamic quantities can be derived,
both with and without angular momentum constraints, analogous the 
expressions given above for the ensemble of ripplon modes.

The same procedure can be used to calculate the density of states without 
the restriction $k_{n,L} \ll 1$\,\AA, if the values of $\nu_{n,m}$
are calculated using the elementary excitation energy of liquid helium
for wavenumber $k_{n,L}$.  However, this no longer results in a density
of states that is a function only of a reduced energy, $E_{\textrm{p}}$,
and thus would have to explicitly calculated for each droplet size.
This has not been pursued in the present work.

\section{Conclusions}

In this work, we have presented simple numerical procedures that allows the
calculation of the density of states of both ripplon and phonon excitations of
a helium nanodroplet as a function of both reduced energy and total angular
momentum quantum numbers.  It is found that these can well approximated by
simple analytical expressions.  It has been found that the distribution of
rotational total angular momentum, for each energy, closely follows that of
rigid spherical top.  The microcanonical expressions for the ripplon density
of states has been used by the author and collaborator for statistical
evaporative cooling calculations of both pure and doped helium nanodroplets,
conserving angular momentum.    That work will be presented in a future
publication.

\section{Acknowledgements}

\begin{acknowledgments}

The author would like to acknowledge the advice and assistance of his
Princeton University collaborations Adriaan Dokter, Roman Schmied, and Giacinto
Scoles.  This work was supported by a grant from the National Science Foundation.

\end{acknowledgments}

\appendix*
\section{Appendix: Reduction of Symmetric Products of the Rotation Group}

In order to compute the density of ripplon states as a function of both
energy and total angular momentum quantum number, one needs to determine the
distribution of states of different angular momentum generated by
$n_{L}$ quanta in each ripplon mode with mode angular momentum quantum
number $L$.  We use the fact that a set of states with total angular 
moment quantum number $L$ and with projection $M = -L, -L+1 \ldots L$
is an irreducible representation of the rotation group of the sphere, K
(or the double group of rotations is half integer angular momentum is
allowed for).  Below, when we speak of a ``state'' with a given total
angular momentum quantum number, $J$, it will be implicit we mean a
degenerate set of $(2J+1)$ eigenstates with the allowed range of projection
quantum numbers.   For the states generated by multiple excitations
in different modes, the total angular momentum distribution is found by
reduction of the direct product representation of the
distributions in each mode.  The reduction of a direct product
representation for the sphere gives the well known triangle rule, 
i.e. products of
states with quantum numbers $J_1$ and $J_2$ give one state each with 
$J = |J_1 - J_2|, |J_1 - J_2| + 1, \ldots, J_1 + J_2$.  The triangle rule can be
applied recursively to determine the total number of states with each total $J$
once we know the distribution of number of states with each $J$ quantum
number for excitation of $n_L$ quanta in a harmonic ripplon mode with angular
momentum quantum number $L$.

The distribution of states produced by multiple excitation in a degenerate
mode is not give by the direct product of that mode with itself $n_L$ times,
but by the states given from the reduction of the symmetric $n_L$ power product
of that mode.   This is the meaning of the widely cited statement that
vibrational quanta are ``Bosons''.  For the point groups relevant to the
spectra of rigid molecules, the reduction of such symmetric products can
be found in a standard text on vibrational spectroscopy~\cite{Wilson55}.
However, the author has not been able to locate equivalent expressions
for the reduction of general symmetric powers of the rotation group. 
In this appendix, we give a recursion relation that has been used to
calculate these up to high values of $L$ and $n_L$.

Each state produced by the symmetric direct product of $n_L$ quanta in mode
with total angular momentum quantum number $L$ can be represented by a set
of $m_{\textrm i}$ quantum numbers (integer or half integer depending
upon $L$) such that 
\begin{equation}
-L \le m_1 \le m_2  \dots \le m_{n_L} \le L
\end{equation}
We will calculate a recursive relationship for $N_M(M,n_L,L)$ which is the
number of $n_L$'th power symmetric product states of mode $L$ such that
the sum of the quantum numbers $m_{\textrm i}$ equals $M$.  For $n_L = 1$,
$N_M(M,1,L) = 1$ if $|M| \le L$ and zero otherwise.
Note that if we restrict $m_{n_L} = L$, then the total number of such
states  that contribute to $N_M(M,n_L,L)$ is equal to $N_M(M-L,n_L-1,L)$.  If
$m_{n_L} = L-1$, then because the rest of the quantum numbers are restricted
to only $2 L$ values, the number of states that contribute to $N_M(M,n_L,L)$
is equal to
$N_M(M- (L-1) + (n-1)/2 ,n_L-1,L-1/2)$.  The last term in the first argument
arises from the ``shift'' in the projection quantum number when mapping the
sum over the $2 n_L$ values to the symmetric direct product for a mode with
angular momentum quantum number $L - 1/2$.  For the general case of
$m_{n_L} = L - k$, the number of such states contributing to $N_M(M,n_L,L)$ is
equal to $N_M(M - (L-k) + k(n_L-1)/2,n_L-1,L-k/2)$.  Taking these
contributions into account as well as that $0 \le k \le 2L$ and that
$N_M(M,n_L,L) = 0$ if $|M| > n_L L$ we can write:
\begin{equation}
N_M(M,n_L,L) = \sum_{k = k_{min}}^{k_{max}} N_M(M - L + k(n_L +1)/2, n_L -1, L -
k/2)
\label{eq:NM}
\end{equation}
Where $k_{min}$ is the maximum value of 0 and $-M+2L-n_{L}L$ and
$k_{max}$ is the minimum value of $2L$ and $L-(M/n_L)$. 
We can then calculate $N_J(J,n_L,L)$, the number of
states with total angular momentum quantum number $J$ by
$N_J(J,n_L,L) = N_M(J,n_L,L) - N_M(J+1,n_L,L)$, as is done in
many introductory Quantum texts when calculating the terms produced by
a given atomic configuration.  It is apparent, since all terms in
Eq.~\ref{eq:NM} shift in their first argument by one when the
first argument on the left shifted by one that exactly the same
recursion relationship can be applied to directly calculate
the $N_J(J,n_L,L)$ values.  However, in such a
calculation, one must replaces $N_J(-J,n_L,L)$
by $-N_J(J-1,n_L,L)$ and $N_J(-1/2,n_L,L) = 0$ when
these negative argument values arise in the recursion.
Also, the recursion is started by using $N_J(J,1,L) = \delta(J,L)$.
It is also easily demonstrated that 
$N_J(J,2,L) = (1/2)\left( 1 + (-1)^{J+2L} \right)$
for $J \le 2L$ (the factor of $2L$ is included to make the
result correct even in the case of half integer quantum numbers).
Note that it is possible to calculate each successive
value of $n_L$ by replacement in a single
stored two dimensional matrix by looping down from the highest $L$
value.

The calculation of the $N_J(J,n_L,L)$ values becomes quite memory demanding
for large values of $n_L,L$.  This is in part because they must be generated
with increasing values of $n_L$, but must be used in the calculation of
the density of states in order of increasing $L$, thus requiring a three
dimensional matrix with $\approx \frac{1}{2} (n_{\textrm{max}}
L_{\textrm{max}})^2$ elements, where $n_{\textrm{max}}$ and $L_{\textrm{max}}$
are the highest values of $n_L$ and $L$ needed.  However, it has been found
that for large $n_L$ and $L$ values the distribution of $N_J(J,n_L,L)$
approaches that of a rotational distribution of a spherical top, i.e. 
\begin{equation}
N_J(J,n_L,L) \approx g(n_L,L) (2J+1) \sqrt{\frac{\beta(n_L,L)^3}{\pi}}
\exp \left( -\beta(n_L,L)(J + 1/2)^2 \right)
\end{equation}
where $g(n_L,L) = \frac{(n_L + 2L)!}{(2L)!n_L!}$ is the total number of states
(including the 2J+1 degeneracies) generated with $J=0 \dots n_L L$.
This functional form can be justified when it is recognized that the central limit
theorem~\cite{Reif65} suggests that $N_M(M,n_L,L)$ should approach a Gaussian
distribution in $M$ for large $n_L$ values and that $N_J(J,n_L,L)$ is the
derivative of $N_M(J,n_L,L)$ in that limit.
We can determine the dependence of $\beta(n_L,L)$ by
equating the ``thermal average'' value of $J(J+1)$ to the exact expression,
Eq.~\ref{eq:J2_nL}, which gives
\begin{equation}
\beta(n_L,L) = \frac{3}{2 n_L \left[ L(L+1) + \frac{1}{2} L (n_L - 1) \right]}
\end{equation}
For $n_L = 14$ and $L \ge 10$, the maximum error in this approximation to
$N_J(J,n_L,L)$ is less than 5\% of the peak value for each $L$.


\newpage

\end{document}